\documentclass[12pt,preprint,showkeywords]{iopart}

\newcommand{\bfm}[1]{\mbox{\boldmath${#1}$}}
\newcommand{\sect}[1]{\setcounter{equation}{0}\section{#1}}

\usepackage{graphicx}
\usepackage{dcolumn}
\usepackage{amssymb}
\usepackage{latexsym}
\begin{document}

\title[Stochastic quantization of interacting classical particles system]
{Stochastic quantization of interacting classical particles system.}

\author{A.M. Scarfone}

\address{Istituto
Nazionale di Fisica della Materia (CNR-INFM) and\\ Departimento di
Fisica - Unit\`{a} del Politecnico di Torino\\ Corso Duca degli
Abruzzi 24, I-10129 Torino, Italy}

\date{\today}
\begin {abstract}
Starting from a many-body classical system governed by a trace-form
entropy we derive, in the stochastic quantization picture, a family
of non linear and non-Hermitian Schr\"odinger equations describing,
in the mean filed approximation, a quantum system of interacting
particles. The time evolution of the main physical observables is
analyzed by means of the Ehrenfest equations showing that, in
general, this family of equations takes into account dissipative and
damped effects due to the interaction of the system with the
background. We explore the presence of steady states by means of
solitons, describing conservative solutions. The results are
specialized to the case of a system governed
by the  Boltzmann-Gibbs entropy. \\

\end {abstract}
\submitto{JSTAT}

\eads{\mailto{antonio.scarfone@polito.it}} \maketitle

\sect{Introduction}

The problem of the quantization of a many-body system attracted the
attention of the physics community immediately in the first years
after the introduction of the quantum mechanics \cite{Fermi}. Until
today, different methods have been introduced to study this problem.
Among the many we recall the Hartree-Fock method, the Thomas-Fermi
approach, the Bohm-Madelung quantization, the second quantization by
Heisenberg, the stochastic quantization by Nelson
and others \cite{Ring}.\\
Concerning a non relativistic system, nonlinear Schr\"odinger
equations (NSEs) are generally employed to take into account,
through the nonlinearity, the interactions among the many particles.
As a paradigm, the NSE with the cubic nonlinearity
\cite{Gross,Pitaevskii} has been widely used in literature to study,
for instance, the Bose-Einstein condensation of alcali atoms like
$^7$Li, $^{23}$Na and $^{87}$Rb \cite{Stringari}.\\
In \cite{Bialynicki}, Bialynicki-Birula and Mycielski (BBM)
introduced the nonlinearity $-b\,\ln(|\psi|^2)\,\psi$ which was
selected from the assumption of a separability condition among
composed systems. Successively, in order to save only partially the
superposition principle, valid in the linear theory, Weinberg
\cite{Weinberg} suggested a very general class of NSEs with an
homogeneous nonlinearity. \\ Motivated by these works, some
experimental tests were carried out based on the neutrons
interferometry \cite{Shimony,Shull} and $^9$Be$^+$ transition
frequency spectroscopy \cite{Bollinger}. All of them fail in
pointing out the existence of nonlinear effects, suggesting in this
way, that there is no real ground, at fundamental level, for such
nonlinearities in the Schr\"odinger equation.\\
Notwithstanding, NSEs can be usefully
applied in the description of extended objects. In fact, depending
on the form of the nonlinearity, some NSEs admit soliton solutions,
i.e. non dispersive wave packets
identifiable with {\em quasi-classical} particles.\\
For instance, in the BBM equation a soliton solution named {\em
gausson} has been studied \cite{Bialynicki,Hasse}. This solution
(\cite{Hefter} and references therein) represents the experimentally
measurable charge or mass density $\rho=|\psi|^2$ of an atomic
nucleus, when it is considered as an elementary object (neglecting,
in this way, its internal quarks structure).\\
Remarkably, the BBM equation is closely connected with the
Boltzmann-Gibbs (BG) entropy. In \cite{Hefter}, it has been shown
that the binding energy released in the splitting of an arbitrary
wave function in non overlapping parts of the same form $\psi({\bfm
x})\to\sum_ip_{_i}^{1/2}\,\psi({\bfm x}-{\bfm x}_{_i})$, with
$p_{_i}$ normalized according to $\sum_i p_{_i}=1$, is given by
$\Delta\,E=b\,S^{\rm BG}(p)$, where $S^{\rm
BG}(p)=-\sum_ip_{_i}\,\ln p_{_i}$. This suggests that the BBM
equation can be used in the thermodynamical description of spatially
extended or self-interacting quantum mechanical objects.\\
Also the BG entropy finds several applications in the study of
quantum systems. For instance, in
\cite{Bialynicki1,Bialynicki2,Maassen} this entropy has been applied
to generalize the Heisenberg uncertainty relation for a pair of
conjugate observables. In \cite{Garbaczewski} a worthwhile
discussion concerning the BG entropy and the Fisher information
measure has been advanced. Author clarifies the role of these
entropies in the Smoluchowski diffusion process which describes the
underlying quantum kinetics of
the linear Schr\"odinger equation,\\
Moreover, in recent years, an intensive research activity has been
focused in the study of complex quantum systems which are intimately
related to generalized versions of the BG entropy
\cite{Tirnakli,Olavo1,Rossani}, although the link between the
microscopic properties of these systems and their statistical
mechanics features is still an open question. In this respect, in
\cite{Naudts} a microscopic derivation of a statistical system
described by a general entropy has been considered. A microcanonical
approach for a quantum system, at equilibrium, is studied in
\cite{Naudts1,Bender} (and references therein) whereas, more
recently in \cite{Scarfone1}, author deals with the canonical
quantization of a classical system governed by an arbitrary entropy
whose kinetics is described by a very general nonlinear
Fokker-Planck equation (NFPE). He obtained a wide family of NSEs
where the structure of the complex nonlinearity is related to the
form of the entropy of the ancestor classical system. The method has
the drawback that, in consequence of the canonical quantization,
this family of NSEs describes the time evolution of isolated quantum
systems conserving energy and momentum.

The purpose of the present paper is to accomplish the quantization
of a many-body system of interacting particles in thermal contact
with a reservoir (the background). This can be performed by using a
generalized version of the stochastic quantization method initially
proposed by Nelson \cite{Nelson1,Nelson2,Nelson3}. In this way, we
derive a family of NSEs with a complex nonlinearity, describing
damped and dissipative processes. We remark that complex
nonlinearities arise whenever the quantum system obeys to a
diffusive continuity equation. This, for example, occurs for the
family of Schr\"odinger equations derived by Doebner and Goldin
\cite{Doebner} as the most general class compatible with the linear
Fokker-Planck equation, or, in \cite{Scarfone0}, where a NSE has
been derived from a generalized Pauli exclusion-inclusion principle
accounted through a suitable modification (in a nonlinear manner) of
the particle quantum current. Also in \cite{Scarfone7}, a class of
NSEs with a complex nonlinearity has been obtained in the stochastic
quantization framework, starting from the most general kinetics
containing a nonlinear drift term and compatible with a linear
diffusion term. Quite generally, all these NSEs can be transformed
in other ones, with a purely real nonlinearity, by means of a gauge
transformation of third kind \cite{Doebner, Scarfone2,Scarfone3}.

Although sometime criticized \cite{Wallstrom}, the Nelson method is,
at present, argument of an intense research activity
\cite{Davison2,Wang,Bacciagaluppi,Petroni}. It merges together both
the Newtonian mechanics and the diffusion process with the prospect
to understand the quantum mechanics in a pure classical language.
Typically, one assumes a quantum system undergoing an irreducible
Brownian motion in the configuration space. Notwithstanding, the
stochastic mechanics is not merely a mathematical description of a
diffusion process. Two main aspects make the difference with the
usual treatment of a standard diffusive process: the prescription
associated to the reverse of the time axis in the diffusion process
and the introduction of a stochastic acceleration, whose definition
has been generalized in several papers \cite{Davison1,Guerra}.

In the following, we
generalize the original Nelson method according to the following two steps:\\
(i) By assuming a nonlinear kinetic underling the stochastic
diffusion motion, so that one can replace the forward and backward
linear Fokker-Planck equations with a couple of nonlinear kinetic
equations, in a manner compatible with the
original NFPE describing the kinetic of a classical system;\\
(ii) By generalizing the definition of the mean acceleration in the
most general quadratic-form, compatible with a local expression of
an {\em Euler} equation.

In the next section 2, we recall the relationship among an arbitrary
entropy and the corresponding NFPE, describing the kinetic of a
classical system. In section 3, we start from this NFPE and derive
the family of NSEs describing a quantum many-body system. In section
4, we investigate the time evolution of the main observables
(Ehrenfest relations) whilst, in section 5, we inquire the existence
of a special class of solutions, the solitons, which describe an
isoentropic and non dissipative process. Finally, in the section 6,
as an example, we discuss the quantization of a classical system
described by the BG entropy. The conclusions are reported in the
final section 7.

\sect{Nonlinear kinetic equation} In literature, there have been
presented many different approach to relate, in a classical system,
the production of entropy to a NFPE \cite{Frank1}.

In the following, we consider a many-body classical system described
with the field $\rho({\bfm x},t)$, labeled by the vector ${\bfm
x}=(x_1,\,x_2,\,\,x_3)$, and normalized in
\begin{equation}
\int\rho({\bfm x},t)\,d{\bfm x}=1 \ .
\end{equation}
Quite generally, the time evolution of this field can by obtained,
in the configuration space, starting from the continuity equation
\begin{equation}
\frac{\partial\rho}{\partial t}+{\bfm\nabla}\cdot{\bfm J}=0 \
,\label{FP}
\end{equation}
and assuming for the current the expression
\begin{equation}
{\bfm J}=-\rho\,{\bfm F}(\rho) \ .
\end{equation}
The thermodynamic force ${\bfm F}(\rho)$, defined through the
relation
\begin{equation}
{\bfm F}(\rho)={\bfm \nabla}\left(\frac{\delta {\mathcal L}}{\delta
\rho}\right) \ ,\label{force}
\end{equation}
is related to the functional
\begin{equation}
{\mathcal L}(\rho)=U-D\,S(\rho) \ ,\label{Lyapunov}
\end{equation}
where
\begin{equation}
U(t)=\int {\mathcal E}({\bfm x})\,\rho({\bfm x},t)\,d{\bfm x} \ ,
\end{equation}
is the mean energy of the system, ${\mathcal E}({\bfm x})$ is the
total energy for particle and $D$ is the constant diffusion
coefficient.\\
Let us assume for the entropy $S(\rho)$ the very general trace-form
expression
\begin{equation}
S(\rho)=-\int d{\bfm x}\int d\rho\,\ln\kappa(\rho) \
,\label{entropy}
\end{equation}
(throughout this paper we use units with the Boltzmann constant
$k_{\rm B}=1$).\\
In equation (\ref{entropy}), the quantity $\kappa(\rho)$ is a smooth
function of the density field $\rho({\bfm x},\,t)$. Depending on the
expression of $\kappa(\rho)$, equation (\ref{entropy}) encompasses
different
entropies already introduced in literature \cite{Kaniadakis}. \\
From (\ref{FP})-(\ref{entropy}) we obtain the following continuity
equation
\begin{equation}
\frac{\partial\rho}{\partial t}+{\bfm\nabla}\cdot\Big[\rho\,{\bfm
v}_{\rm drift}-D\,f_0(\rho)\,{\bfm \nabla}\rho\Big]=0 \ ,\label{FP1}
\end{equation}
where we have introduced the drift velocity
\begin{equation}
{\bfm v}_{\rm drift}=-{\bfm\nabla}\,{\mathcal E}({\bfm x}) \
,\label{lc}
\end{equation}
whilst the function $f_0(\rho)$ is given by
\begin{equation}
f_0(\rho)=\rho\,\frac{\partial}{\partial\rho}\ln\kappa(\rho) \ .
\end{equation}
Equation (\ref{FP1}) is a NFPE in the Smoluchowski picture since it
describes the kinetics process in the configuration space rather
than
in the momentum space.\\
We observe that in equation (\ref{FP1}) the total current ${\bfm
J}={\bfm J}_{\rm drift}+{\bfm J}_{\rm diff}$ is the sum of the
standard linear drift current ${\bfm J}_{\rm drift}=\rho\,{\bfm
v}_{\rm drift}$ and the nonlinear diffusion current ${\bfm J}_{\rm
diff}=-D\,f_0(\rho)\,{\bfm\nabla}\rho$ which reduces to
the standard Fick form ${\bfm J}_{\rm Fick}=-D\,{\bfm\nabla}\rho$
when
$\kappa(\rho)=\rho$.\\
Since $D$ is constant, equation (\ref{FP1}) describes an isothermal
diffusive process where the system is in thermal contact with a
reservoir. In fact, it is known that for an isothermal process the
diffusion coefficient plays the role of the inverse of a kind of
temperature \cite{Frank1}. In this case, we easily recognize that
equation (\ref{Lyapunov}) defines a free energy for the system we
are looking at. Actually , we can easily verify that the quantity
${\mathcal L}(\rho)$ is a Lyapunov function for the given problem
since, according to equation (\ref{FP1}), we have
\begin{equation}
\frac{d\,{\mathcal
L}}{d\,t}=-\int\rho\,\left[{\bfm\nabla}\,{\mathcal E}({\bfm
x})-D\,{\bfm\nabla}\left(\frac{\delta\,S}{\delta\,\rho}\right)\right]^2\,d{\bfm
x}\leq0 \ ,
\end{equation}
where equality holds at equilibrium.

It is worthy to recall that recently, different NFPEs have been
related with generalized entropies of the kind (\ref{entropy}) by
using several methods substantially equivalent to the one described
above. In \cite{Chavanis1}, generalized Kramers and Smoluchowski
equations were derived by means of a variational principle, which
minimizes the dissipative ratio of a generalized free energy
equivalent to equation (\ref{Lyapunov}). The derivation of a
Smoluchowski equation starting from a Kramers equation has been
rigorously derived in \cite{Chavanis2} by mean of the Chapman-Enskog
expansion in the high friction limit. Finally, in \cite{Kaniadakis},
a kinetic interaction principle has been proposed to derive a family
of NFPEs starting from a classical Markovian process describing the
kinetics of a system of $N$-body particles. The kinetic interaction
principle is realized through a suitable factorization of the
transition probability $\pi(t,\,{\bfm x}\to{\bfm y})=r(t,\,{\bfm
x},\,{\bfm y})\,a(\rho)\,b(\rho')\,c(\rho,\,\rho')$ in terms of the
population of the initial site $\bfm x$ and the final site $\bfm y$,
with $\rho=\rho(t,\,{\bfm x})$ and $\rho'=\rho(t,\,{\bfm y})$, where
the functions $a(\rho)$, $b(\rho')$ and $c(\rho,\,\rho')$ fix the
nonlinearities in the NFPE and imposes the form of the entropy
associated to the system.

\sect{Stochastic quantization of the many-body system}

In the stochastic quantization framework, particles are assumed
undergoing a time-asymmetric Markovian random process. The space
coordinate $\bfm x$ satisfies a couple of stochastic differential
equations
\begin{equation}
d{\bfm x}(t)={\bfm v}^{(\pm)}({\bfm x},t)\,dt+d{\bfm W}^{(\pm)}(t) \
,\label{lang}
\end{equation}
describing a forward $(+)$ and backward $(-)$ process, respectively.
They can be seen as a system of Langevin equations in the
configuration space, where ${\bfm v}^{(\pm)}({\bfm x},\,t)$ are the
forward and backward velocities, two vector valued functions
depending on space and time, whilst the functions ${\bfm
W}^{(\pm)}(t)$ describe a Wiener process.\\
According for equation (\ref{lang}), ${\bfm x}(t)$ is not a
differentiable function. This means that there is no velocity as
standard time derivative. Thus, one introduces a forward and a
backward time derivative given by
\begin{equation}
\left({d\over
d\,t}\right)^{(\pm)}=\frac{\partial}{\partial\,t}+{\bfm
v}^{(\pm)}\cdot{\bfm\nabla}\pm D\,\Delta \ ,\label{dbf}
\end{equation}
so that
\begin{equation}
{\bfm v}^{(\pm)}({\bfm x},t)=\left({d\over dt}\right)^{(\pm)}{\bfm x}(t) \ .
\end{equation}
Let us identify the field $\rho({\bfm x},t)$ with the probability
density of a quantum system at the position ${\bfm x}(t)$. We assume
that it fulfils the following forward and backward NFPEs
\begin{eqnarray}
\frac{\partial\rho}{\partial t}+{\bfm\nabla}\cdot\left[{\bfm
v}^{(\pm)}\,\rho\mp D\, f^{(\pm)}(\rho){\bfm\nabla}\rho\right]=0 \
,\label{fbnfp}
\end{eqnarray}
where, without lost of generality, we pose
\begin{equation}
f^{(\pm)}(\rho)=2\,\rho\,\frac{\partial}{\partial\rho}\ln\kappa^{(\pm)}(\rho)
\ ,
\end{equation}
with $\kappa^{(\pm)}(\rho)$ two arbitrary functions. Clearly, the
form of equations (\ref{fbnfp}) is reminiscent of the expression
(\ref{FP1}) which describes the kinetic of a classical system.

From equations (\ref{fbnfp}), we see that forward and backward
velocities are related each other in
\begin{equation}
{\bfm v}^{(+)}={\bfm v}^{(-)}+2\,D\,{\bfm\nabla} g(\rho) \ ,
\end{equation}
where
\begin{equation}
g(\rho)=\ln \Big(\kappa^{(+)}(\rho)\,\kappa^{(-)}(\rho)\Big) \
.\label{gg}
\end{equation}
As customary, we introduce the mean velocity and the osmotic
velocity according to the relations
\begin{equation}
{\bfm v}={1\over2}\Big({\bfm v}^{(+)}+{\bfm v}^{(-)}\Big) \
,\hspace{10mm}{\bfm u}={1\over2}\,\Big({\bfm v}^{(+)}-{\bfm
v}^{(-)}\Big) \ .\label{vu}
\end{equation}
The former quantity represents the global velocity of the density
shape of the quantum fluid, whilst the later, which takes the
expression
\begin{equation}
{\bfm u}=D\,{\bfm\nabla}g(\rho) \ ,\label{upm}
\end{equation}
has an intrinsic stochastic origin and is related to the spatial
variation
of the density. \\
Endowed with the definitions (\ref{vu}), by averaging equations
(\ref{fbnfp}), we obtain
\begin{equation}
\frac{\partial\rho}{\partial t}+{\bfm\nabla}\cdot \Big[\rho\,{\bfm
v}-D\,f_0(\rho)\,{\bfm\nabla}\rho\Big]=0 \ ,\label{cee1}
\end{equation}
a nonlinear continuity equation for the field $\rho({\bfm x},t)$,
where
\begin{equation}
f_0(\rho)={1\over2}\,\Big[f^{(+)}(\rho)-f^{(-)}(\rho)\Big]=
\rho\,\frac{\partial}{\partial\rho}\ln \kappa(\rho) \ ,\label{f0}
\end{equation}
and
\begin{equation}
\kappa(\rho)={\kappa^{(+)}(\rho)\over\kappa^{(-)}(\rho)} \
.\label{k}
\end{equation}
Equation (\ref{cee1}) is formally equivalent to the NFPE
(\ref{FP1}). According to this correspondence, the expression of the
function $\kappa(\rho)$ is fixed through the entropy of the ancestor
classical system although the same entropy does not determine
univocally the functions $\kappa^{(\pm)}(\rho)$. From a mathematical
point of view, this means that there are infinitely possible choices
for the functions $\kappa^{(\pm)}(\rho)$, all compatible with the
form of the entropy (\ref{entropy}), according to equation
(\ref{k}). Note that equation (\ref{cee1}) is consistent with the
kinetic interaction principle proposed in \cite{Kaniadakis}, where a
very general NFPE, whose equation (\ref{cee1}) is a particular case
with the linear drift, has been derived starting from a master Pauli
equation describing a Markovian process.

Let us now introduce the stochastic acceleration. In the original
Nelson work it is defined in
\begin{equation}
{\bfm a}={1\over2}\,\left[\left({d\over dt}\right)^{(+)}{\bfm
v}^{(-)}+\left({d\over dt}\right)^{(-)}{\bfm v}^{(+)}\right] \ ,
\end{equation}
i.e. by means of the algebraic mean of the quadratic operators
$(d/dt)^{(+)}(d/dt)^{(-)}$ and $(d/dt)^{(-)}(d/dt)^{(+)}$ acting on
the vector ${\bfm x}(t)$. The non uniqueness of the definition of
${\bfm a}$ has been noted in \cite{Davison1} and alternative
definitions have been proposed in \cite{Davison1,Guerra}.
Hereinafter, we assume the most general quadratic form, in the
forward and backward time derivatives, compatible with a local
expression for an {\em Euler} equation describing the dynamics of
${\bfm v}$. Thus, we pose
\begin{equation}
\hspace{-20mm}{\bfm a}=\tau\,\left(\frac{d}{dt}\right)^{(+)}{\bfm
v}^{(+)}+\mu\,\left[\left(\frac{d}{dt}\right)^{(+)}{\bfm v}^{(-)}+
\left(\frac{d}{dt}\right)^{(-)}{\bfm v}^{(+)}\right]
+\nu\,\left(\frac{d}{dt}\right)^{(-)}{\bfm v}^{(-)} \ , \label{acc}
\end{equation}
with the condition $\tau+2\,\mu+\nu=1$.\\
We observe that the original Nelson definition is recovered from
equation (\ref{acc}) by posing $\tau=\nu=0$ and $\mu=1/2$ whilst,
the definition advanced in \cite{Davison1} is obtained by posing
$\tau=\nu=1/8$ and $\mu=3/8$. Moreover, in the framework of a
stochastic Lagrangian formulation \cite{Guerra}, compatible with the
variational stochastic calculus introduced in \cite{Yasue}, another
definition of ${\bfm a}$ has been advanced. It follows from equation
(\ref{acc}) by posing $\tau=\nu=1/2$ and $\mu=0$.

As customary, let us introduce the potential field $\Sigma({\bfm
x},t)$ related to the mean velocity through the relation
\begin{equation}
m\,{\bfm v}={\bfm\nabla}\Sigma \ .\label{ls}
\end{equation}
Remarkably, this equation is formally equivalent to the relation
(\ref{lc}) among the classical drift velocity and the particle
energy spectrum although the fields $\Sigma({\bfm x},t)$ and
${\mathcal E}({\bfm x})$ represent two different physical
quantities. In this sense, the quantum mean velocity $\bfm v$
assumes a different meaning, in the quantum picture, with respect to
the meaning that the classical drift velocity ${\bfm v}_{\rm drift}$
has
in the classical picture.\\
Taking into account that the both fields $\bfm v$ and $\bfm u$ are
given through the gradient of the scalar functions $g(\rho)$ and
$\Sigma({\bfm x},t)$, respectively, and recalling the Newton's law,
from the definition (\ref{acc}) one obtain
\begin{eqnarray}
\nonumber \frac{\partial{\bfm v}}{\partial t}&+&{\bfm
v}\cdot{\bfm\nabla}{\bfm v}+a_1\,\frac{\partial{\bfm u}}{\partial
t}+a_1\,{\bfm\nabla}({\bfm u}\cdot{\bfm v})+D\,a_1\,\Delta{\bfm
v}\\&+&a_2\,{\bfm u}\cdot{\bfm\nabla}{\bfm u}+D\,a_2\,\Delta{\bfm
u}+{1\over m}\,{\bfm\nabla}V=0 \ ,\label{vt}
\end{eqnarray}
where $a_1=\tau-\nu$ and $a_2=\tau-2\,\mu+\nu$.\\ In effects,
(\ref{vt}) is an {\em Euler} equation for the mean velocity field
${\bfm v}({\bfm x},t)$. We remark that, since the osmotic velocity
is a function of the field $\rho({\bfm x},t)$, the quantity
$\partial{\bfm u}/\partial t$ can be easily handled by
using the continuity equation (\ref{cee1}).\\
By employing the definitions (\ref{upm}) and (\ref{ls}), the {\em
Euler} equation takes the expression
\begin{equation}
\frac{\partial\Sigma}{\partial
t}+\frac{({\bfm\nabla}\Sigma)^2}{2\,m}+f_1(\rho)\,\Delta\Sigma+
f_2(\rho)\,({\bfm\nabla}\rho)^2+f_3(\rho)\,\Delta\,\rho+V=0 \
,\label{eu}
\end{equation}
where
\begin{eqnarray}
\nonumber
&&f_1(\rho)=a_1\,D\,\left(1-\rho\,\frac{\partial g(\rho)}{\partial\rho}\right) \ ,\\
&&f_2(\rho)=m\,D^2\,\left[a_1\,\frac{\partial
f_0(\rho)}{\partial\rho}\,\frac{\partial g(\rho)}{\partial\rho}
+{a_2\over2}\left(\frac{\partial
g(\rho)}{\partial\rho}\right)^2+a_2\,\frac{\partial^2
g(\rho)}{\partial\rho^2}\right] \
,\label{f}\\
\nonumber
&&f_3(\rho)=m\,D^2\,\Bigg(a_1\,f_0(\rho)+a_2\Bigg)\,\frac{\partial
g(\rho)}{\partial\rho} \ .
\end{eqnarray}
As known, in the hydrodynamic representation of a quantum system
\cite{Bohm,Madelung}, the quantum effects are take into account
through the introduction of the quantum potential $U_{\rm
q}(\rho)=-(\hbar^2/2\,m)\,\Delta\sqrt{\rho}/\sqrt{\rho}$. In the
original Nelson formulation, the potential $U_{\rm q}(\rho)$
arises from the particular expression assumed by the osmotic
velocity that, in that case, is given by
\begin{equation}
{\bfm u}=D\,\ln\rho \ .\label{un}
\end{equation}
In the nonlinear case studied in this paper, the same form
(\ref{un}) can be recovered by posing
\begin{equation}
\kappa^{(+)}(\rho)=\sqrt{\rho\,\kappa(\rho)} \
,\hspace{10mm}\kappa^{(-)}(\rho)=\sqrt{\rho\over\kappa(\rho)} \
,\label{k+-}
\end{equation}
although, according with the non uniqueness of
$\kappa^{(\pm)}(\rho)$, the choice (\ref{k+-}) is not unique.\\In
this way, equation (\ref{eu}) simplifies in
\begin{equation}
\frac{\partial\Sigma}{\partial
t}+\frac{({\bfm\nabla}\Sigma)^2}{2\,m}+U_{\rm q}(\rho) +W(\rho)+V=0
\ ,\label{hj}
\end{equation}
where
\begin{equation}
W(\rho)=m\,D^2\,{a_1\over\rho}\,{\bfm\nabla}\cdot\Big(f_0(\rho){\bfm\nabla}\rho\Big)
\ ,
\end{equation}
and $a_2=-(\hbar/2\,m\,D)^2$.\\
Equations (\ref{cee1}) and (\ref{hj}) govern the time evolution of
the fields $\rho({\bfm x},t)$ and $\Sigma({\bfm x},t)$. Actually,
(\ref{cee1}) is a quantum continuity equation for the field
$\rho({\bfm x},t)$ describing the kinetics of the quantum fluid {\em
a l\`{a}} Bohm-Madelung, where the quantum current, according to the
position (\ref{ls}), assumes the expression
\begin{equation}
{\bfm j}=\rho\,{{\bfm \nabla}\Sigma\over
m}-D\,f_0(\rho)\,{\bfm\nabla}\rho \ .
\end{equation}
Differently, (\ref{hj}) can be interpreted as a nonlinear {\em
Hamilton-Jacobi} equation for the potential field $\Sigma({\bfm
x},t)$ whose dynamics is ruled through the quantum potential, the
extra nonlinear potential $W(\rho)$ and the external potential $V$.

Finally, with the position
\begin{equation}
\psi({\bfm x},\,t)=\sqrt{\rho({\bfm
x},\,t)}\,\exp\left(\frac{i}{\hbar}\,\Sigma({\bfm x},\,t)\right) \
,\label{ans}
\end{equation}
both equations (\ref{cee1}) and (\ref{hj}) can be joined to
obtain the following NSE
\begin{equation}
i\,\hbar\,\frac{\partial\psi}{\partial t}=\mathrm{H}\,\psi \
,\label{se}
\end{equation}
where $\mathrm{H}$ is the non-Hermitian operator
\begin{equation}
\mathrm{H}=-\frac{\hbar^2}{2\,m} \,\Delta+\Lambda(\rho)+V \ ,
\end{equation}
with a nonlinearity $\Lambda(\rho)$ depending only on the field
$\rho({\bfm x},t)$. It is given by
\begin{equation}
\Lambda(\rho)={\lambda\over\rho}\,{\bfm\nabla}\cdot\Big(f_0(\rho){\bfm\nabla}\rho\Big)
\ ,\label{nl}
\end{equation}
where the complex coupling constant takes the expression
\begin{equation}
\lambda=m\,D^2\,a_1+i\,{\hbar\over2}\,D \ .\label{la}
\end{equation}
The family of NSEs (\ref{se}) describes, in the mean filed
approximation, the dynamics of a quantum system of identical
particles interacting with a thermal reservoir. The nonlinearity
$\Lambda(\rho)$ is fixed through the entropy (\ref{entropy}) which
governs the kinetics of the ancestor classical system and
consequently determines also the continuity equation of
the quantum system.\\

In conclusion, let us make a few considerations:

- First, the original Nelson derivation is recovered by posing
$\tau=\nu=0$ and $\mu=1/2$. This imply $a_2=-1$ and we recover the
well know relation $D=\hbar/2\,m$ among the diffusion coefficient
and the mass of the particles of the system \cite{Nelson1}.
Moreover, according to the original proposal, we must
pose $\kappa^{(+)}(\rho)=\kappa^{(-)}(\rho)=\sqrt{\rho}$. As a
consequence $f_0(\rho)$ vanishes and equation (\ref{se}) reduces to
the ordinary linear Schr\"odinger equation.

- Second, when $a_1$ vanishes, equation (\ref{se}) describes a family
of NSEs with a purely imaginary nonlinearity. In this case only the
continuity equation acquires an extra nonlinear term (the diffusive
one) whilst the {\em Euler} equation maintains the same expression of
the Bohm-Madelung theory.

- Finally, by means of a nonlinear gauge transformation
\cite{Doebner,Scarfone2,Scarfone3}, the family of NSEs (\ref{se})
can be
transformed in another one with a purely real nonlinearity.\\
In fact, by performing the transformation
\begin{equation}
\psi({\bfm x},t)\to\phi({\bfm x},t)=\psi({\bfm
x},t)\,\exp\left(-{i\over\hbar}\,m\,D\,\int{f_0(\rho)\over\rho}\,d\rho\right)
\ ,\label{gauge}
\end{equation}
the nonlinearity (\ref{nl}) changes according to
\begin{equation}
\Lambda(\rho)\to\widetilde\Lambda(\rho,\,\sigma)=g_1(\rho)\,\Delta\sigma
+g_2(\rho)\left({\bfm\nabla}\rho\right)^2+ g_3(\rho)\,\Delta\rho \
,\label{nl1}
\end{equation}
with
\begin{eqnarray}
\nonumber &&g_1(\rho)=-D\,f_0(\rho) \
,\\
&&g_2(\rho)=m\,D^2\left[{1\over2}\,\left({f_0(\rho)\over\rho}\right)^2+{a_1\over\rho}\,{\partial
f_0(\rho)\over\partial\rho}\right] \
,\\
\nonumber &&g_3(\rho)=m\,D^2\,a_1\,{f_0(\rho)\over\rho} \ .
\end{eqnarray}
In equation (\ref{nl1}), $\sigma({\bfm x},t)$ is the phase of the
new field $\phi({\bfm x},t)$, related to the phase of the old field
$\psi({\bfm x},t)$ through the relation
\begin{equation}
\sigma({\bfm x},t)=\Sigma({\bfm
x},t)-m\,D\,\int{f_0(\rho)\over\rho}\,d\rho \ .
\end{equation}
We observe that the new nonlinearity
$\widetilde\Lambda(\rho,\,\sigma)$, albeit purely real, now depends
through the both fields $\rho({\bfm x},t)$ and $\sigma({\bfm x},t)$.

\sect{Time evolution of observables}

The family of NSEs (\ref{se}) describes dissipative and damped
processes. To show this we derive the
Ehrenfest relations for the main observables of the system.\\
In passing, we notice that, as a consequence of the continuity
equation (\ref{cee1}), the normalization of the wave function
\begin{equation}
\int \rho({\bfm x},t)\,d{\bfm x}=1 \ ,\label{nor}
\end{equation}
is preserved in time. We recall that nonlinear evolution equations,
in general, are not ray-invariants. This means that, given a
solution $\psi$ of equation (\ref{se}), $A\,\psi$ with $A$ a
constant, does not fulfill the same equation. Therefore, in this
case the normalization condition (\ref{nor}) can be accomplished by
fixing appropriately one of the free parameters of the model. As
discussed in \cite{Weinberg}, the ray-invariance can be restored
when equation (\ref{se}) is homogeneous in $\rho({\bfm x},t)$.

Let us now introduce the linear momentum ${\bfm
P}=\int\rho\,{\bfm\nabla}\Sigma\,d{\bfm x}$. By using equations
(\ref{cee1}) and (\ref{hj}), we can derive the following continuity
equation
\begin{equation}
\frac{\partial{\bfm p}}{\partial t}+{\bfm\nabla}\cdot{\mathcal
T}=\bfm{\mathcal R}_{\rm P} \ ,\label{tp}
\end{equation}
for the density  ${\bfm p}=\rho\,{\bfm \nabla}\Sigma$, where the
momentum-stress tensor ${\mathcal T}$ has entries
\begin{equation}
\nonumber {\mathcal T}_{ij}={\rho\over
m}\,\partial_i\Sigma\,\partial_j\Sigma+\frac{\hbar^2}{2\,m\,\rho}\,\partial_i\rho\,\partial_j\rho
+\delta_{ij}\,G_1(\rho) \ ,
\end{equation}
with
\begin{equation}
G_1(\rho)=-{\hbar^2\over4\,m}\,\Delta\rho+a_1\,m\,D^2\int\rho\,{d\over
d\rho}\left[{1\over\rho}\,{\bfm\nabla}\cdot\Big(f_0(\rho)\,{\bfm\nabla}\rho\Big)\right]\,d\rho
\ .
\end{equation}
Clearly, the momentum source term $\bfm{\mathcal R}_{\rm p}$ is
responsible of the non conservation of ${\bfm P}$. Its expression is
given by
\begin{equation}
\bfm{\mathcal R}_{\rm P}=G_2(\rho)\,{\bfm\nabla}\Sigma-\rho\,{\bfm
\nabla}V \ ,\label{rp}
\end{equation}
with
\begin{equation}
G_2(\rho)=D\,{\bfm\nabla}\cdot\Big(f_0(\rho)\,{\bfm\nabla}\rho\Big)
\ .
\end{equation}

In the same way, by defining the energy of the system
$E=-\int\rho\,(\partial\Sigma/\partial t)\,d{\bfm x}$, we can derive
the evolution equation for the density
$\epsilon=-\rho\,\partial\Sigma/\partial t$, which reads
\begin{equation}
\frac{\partial \epsilon}{\partial t}+{\bfm\nabla}\cdot{\bfm J}_{\rm
E}=\rho\,{\mathcal R}_{\rm E} \ ,\label{te}
\end{equation}
where the energy current ${\bfm J}_{\rm E}$ is given by
\begin{equation}
{\bfm J}_{\rm E}=-{\rho\over
m}\,{\bfm\nabla}\Sigma\,\frac{\partial\Sigma}{\partial
t}-{\hbar^2\over4\,m}{{\bfm\nabla}\rho\over\rho}\,\frac{\partial\rho}{\partial
t} \ .
\end{equation}
The energy source term takes the form
\begin{equation}
{\mathcal R}_{\rm E}=-G_2(\rho)\,\frac{\partial\Sigma}{\partial
t}+G_3(\rho) \,\frac{\partial\rho}{\partial t}+\rho\,\frac{\partial
V}{\partial t} \ ,\label{re}
\end{equation}
with
\begin{equation}
G_3(\rho)=a_1\,m\,D^2\,\rho\,\frac{\partial}{\partial\rho}
\left[{1\over\rho}\,{\bfm\nabla}\cdot\left(\frac{f_0(\rho)}{\rho}\,{\bfm\nabla}\rho\right)
\right] \ ,
\end{equation}
and the quantities $\partial\rho/\partial t$ and
$\partial\Sigma/\partial t$ are computed through equations
(\ref{cee1}) and (\ref{hj}), respectively.

By integrating equations (\ref{tp}) and (\ref{te}) on the whole
configuration space and assuming uniform boundary conditions on the
field $\rho({\bfm x},t)$ we obtain the relations
\begin{equation}
\frac{d{\bfm P}}{dt}=\Big\langle{\bfm{\mathcal R}_{\rm
P}}\Big\rangle \ ,\hspace{28mm}\frac{dE}{dt}=\Big\langle{\mathcal
R}_{\rm E}\Big\rangle \ ,\label{1}
\end{equation}
where $\langle{\mathcal R}_{\rm x}\rangle=\int{\mathcal R}_{\rm
x}\,d{\bfm x}$.\\ A further couple of relations easily derivable
from equations (\ref{tp}) and (\ref{cee1}) is given by
\begin{equation}
\frac{d{\bfm L}}{dt}=\Big\langle{\bfm x}\times\bfm{\mathcal R}_{\rm
P}\Big\rangle \ ,\hspace{20mm} \frac{d{\bfm x}_{\rm
mc}}{dt}=\frac{\bfm P}{m} \ ,\label{4}
\end{equation}
with ${\bfm L}=\int{\bfm x}\times{\bfm p}\,d{\bfm x}$ the total
angular momentum and ${\bfm x}_{mc}=\int\rho\,{\bfm x}\,d{\bfm x}$
the mass center of the system.

Thus, as stated by relations (\ref{1})-(\ref{4}), equation
(\ref{se}) describes a quantum dissipative process due to the
presence of the sources ${\mathcal R}_{\rm P}$ and ${\mathcal
R}_{\rm E}$. Notwithstanding, depending on the form of $f_0(\rho)$,
there could exist special classes of solutions describing
conservative systems.
\sect{Soliton solutions}\label{free}

Among the many solutions of a NSE, solitons are of particular
interest in Physics. As known, soliton solutions arise when the
dispersive effects, due to the term $\Delta\psi$ are balanced by the
nonlinear term $\Lambda(\rho)$. In this case, the many particle
system moves coherently giving origin to an extended object with a
particle-like behavior preserving its shape in time.

When the system is constrained in one dimensional space, without the
external potential $V=0$, a soliton solution can be obtained through
the following ansatz
\begin{equation}
\psi(x,t)=\sqrt{\rho(\xi)}\,\exp\left[{i\over\hbar}
\Big(s(\xi)-{1\over2}\,m\,v^2\,t\Big)\right] \ ,
\end{equation}
where $\xi=x-v\,t$ and $v$ is the soliton velocity.\\
Accounting for the relations $\partial/\partial t=-v\,d/d\xi$ and
$\partial/\partial x=d/d\xi$, equations (\ref{cee1}) and (\ref{hj})
form a couple of ordinary differential equations for the scalar
fields $s(\xi)$ and $\rho(\xi)$
\begin{eqnarray}
&&\hspace{-10mm}\rho\,\frac{d
s}{d\xi}-m\,D\,f_0(\rho)\,\frac{d\rho}{d\xi}-m\,v\,\rho=0
\ ,\label{s1}\\
&&\hspace{-10mm}v\,\frac{d
s}{d\xi}-\frac{1}{2\,m}\left(\frac{ds}{d\xi}\right)^2-U_{\rm
q}(\rho)
-a_1\,m\,D^2\,{1\over\rho}\frac{d}{d\xi}\left(f_0(\rho)\,\frac{d\rho}{d\xi}\right)
+{1\over2}\,m\,v^2=0 \ .\label{s2}
\end{eqnarray}
By introducing the function
\begin{equation}
y(\rho)=\left({1\over\rho}\,\frac{d\rho}{d\xi}\right)^2 \
,\label{so}
\end{equation}
we can obtain the following first order differential equation
\begin{equation}
\frac{dy(\rho)}{d\rho}+F_1(\rho)\,y(\rho)+F_2(\rho)=0 \ ,\label{de}
\end{equation}
where
\begin{eqnarray}
&&F_1(\rho)={1\over\rho}{\Big[f_0(\rho)\,(f_0(\rho)+2\,a_1)-a_0
+2\,a_1\,\rho\,(df_0(\rho)/d\rho)\Big]\over a_1\,f_0(\rho)-a_0} \
,\\
&&F_2(\rho)={1\over\rho}\frac{2\,\left(v/D\right)^2}{a_1\,f_0(\rho)-a_0}
\ ,
\end{eqnarray}
and $a_0=(\hbar/2\,m\,D)^2$.\\
Equation (\ref{de}) is solved by quadrature so that, accounting for
equation (\ref{so}), we get
\begin{equation}
\xi=\int{1\over\rho}\left[C\,e^{-\int F_1(\rho')\,d\rho'}-\int
F_2(\rho')\,e^{\int
F_1(\rho'')\,d\rho''}\,d\rho'\right]^{-1/2}\,d\rho \ ,\label{shape}
\end{equation}
where $C$ is the integration constant.\\ This equation defines
implicitly the shape of the soliton. Physically relevant solutions
arise when $\rho(\xi)$ vanishes in a sufficiently fast way for
$\xi\to\pm\infty$, so that the integral
$\int_{-\infty}^{+\infty}\rho(\xi)\,d\xi$ converges. Clearly, this
last condition selects the form of $f_0(\rho)$ and consequently the
possible NSEs admitting soliton solutions.\\ We remark that this
special class of solutions is not dissipative nor damped. They are
steady states for the system under inspection. In fact, we easily
verify that
\begin{equation}
P=\int\limits_{-\infty}\limits^{+\infty}\rho\,\frac{d\Sigma}{dx}\,dx
=\int\limits_{-\infty}\limits^{+\infty}\rho\,\frac{ds}{d\xi}\,d\xi \
,
\end{equation}
and according to the boundary condition $\rho(\pm\infty,t)\to0$ we
obtain that
\begin{equation}
\frac{\partial P}{\partial t}=-v\,\frac{dP}{d\xi}=0 \ .
\end{equation}
In the same way
\begin{equation}
E=\int\limits_{-\infty}\limits^{+\infty}\rho\,\frac{d\Sigma}{dt}\,dx
=-v\,\int\limits_{-\infty}\limits^{+\infty}\rho\,\frac{ds}{d\xi}
\,d\xi-{1\over2}\,m\,v^2 \ ,
\end{equation}
so that
\begin{equation}
\frac{\partial E}{\partial t}=-v\,\frac{dE}{d\xi}=0 \ .
\end{equation}
Actually, the solution (\ref{shape}) describes a particle-like
object traveling with constant velocity $v$, momentum $P=m\,v$ and
energy $E=m\,v^2/2$.  It is also easily to verify that this solution
describes an isoentropic process. This is consistent with the
meaning of the entropy as a measure of the information. In fact,
since the soliton does not change its shape in time, it is
reasonable that the information (entropy) contained in it does not
change in time, accordingly within the relation $dS(\rho)/dt=0$.

\sect{An example}\label{example}

As an example, let us consider the quantization of a classical
system described by the Boltzmann-Gibbs entropy
\begin{equation}
S^{\rm BG}(\rho)=-\int\rho\,\ln\rho\,d{\bfm x} \ ,
\end{equation}
which is obtainable from equation (\ref{entropy}) by posing
$\kappa(\rho)=\rho/e$. This implies $f_0(\rho)=1$, so that the
kinetics of the classical system is governed by the linear
Fokker-Planck equation.\\
The quantum system is described by the following homogeneous NSE
\begin{equation}
i\,\hbar\,\frac{\partial\psi}{\partial t}=-\frac{\hbar^2}{2\,m}
\,\Delta\psi+\lambda\,{\Delta\rho\over\rho}\,\psi+V\,\psi \
,\label{se1}
\end{equation}
where $\lambda$ is given in equation (\ref{la}). This evolution
equation belongs to the family of NSEs study in \cite{Doebner} and
derived on the physical ground as the most general class of
nonlinear quantum evolution equations compatible with the linear
Fokker-Planck
equation.  \\
By employing the gauge transformation (\ref{gauge}), equation
(\ref{se1}) reduces to a NSE with a purely real nonlinearity
\begin{equation}
i\,k\,\frac{\partial\phi}{\partial
t}=-\frac{k^2}{2\,m}\,\Delta\phi+W(\rho,\sigma)\,\phi+V\,\phi \ ,
\end{equation}
where
\begin{equation}
W(\rho,\sigma)=-D\,\Delta\sigma+{1\over2}\,m\,D^2\left({{\bfm
\nabla}\rho\over\rho}\right)^2+a_1\,m\,D^2\,{\Delta\rho\over\rho} \
,
\end{equation}
and
\begin{equation}
\sigma=\Sigma-m\,D\,\ln\rho \ .
\end{equation}
 By posing $a_1=(a_0-1)/2$ and $V=0$, equation (\ref{se1}) admits
a physical solution whose density and phase read
\begin{eqnarray}
&&\rho({\bfm x},t)={A\over
t^3}\,\exp\left(-{1\over4}\,m\,c_1\,\left({{\bfm
x}\over t}\right)^2\right) \ ,\label{sol1}\\
&&\Sigma({\bfm x},t)=c_2+{1\over
t}\,\left[{3\,c_1\over16}\,(4\,m^2\,D^2+\hbar^2)+{1\over2}\,m\,{\bfm
x}^2\right]-{c_1\over4}\,m^2\,D\left({{\bfm x}\over t}\right)^2 \
,\label{sol2}
\end{eqnarray}
respectively, where $A,\,c_1>0$ and $c_2$ are constants. Accounting for the
homogeneity of equation (\ref{se1}) we can accomplish the
normalization of $\rho({\bfm x},t)$ by posing $A=(m\,c_1/4\,\pi)^{3/2}$.\\
Solution (\ref{sol1})-(\ref{sol2}) describes a dispersive wave
function spreading out for $t\to+\infty$, since $\rho({\bfm
x},+\infty)\to0$. Correspondingly, we have also ${\bfm
P}(+\infty)\to0$ and $E(+\infty)\to0$, indicating that this solution
describes a not reversible process, damped and dissipative. It is
also possible to verify that the corresponding entropy is given by
\begin{equation}
S^{\rm BG}=\ln t+C\,t-\ln A \ ,
\end{equation}
with $C=12\,(\pi/m\,c_1)^{3/2}$, which grows in time.

With the same choice for the constant $a_1$ and for the potential
$V$, equation (\ref{se1}) admits a soliton solution. Its shape has
the form
\begin{equation}
\rho(\xi)=A\,e^{-{\alpha\over4}\,\xi^2} \ ,\label{gausson}
\end{equation}
where $\alpha=(4\,m\,v)^2/(4\,m^2\,D^2+\hbar^2)$ and
$A=\sqrt{\alpha/4\,\pi}$. The phase of the wave function is given by
equation (\ref{s1}) and takes the expression
\begin{equation}
\Sigma(\xi,t)=-{1\over4}\,m\,D\,\alpha\,\xi^2+m\,v\,\xi-{1\over2}\,m\,v^2\,t
\ .\label{fgausson}
\end{equation}
The Gaussian shape (\ref{gausson}) is known in literature as a {\em
gausson} \cite{Bialynicki,Hasse} and it was derived firstly as the
soliton solution of the BBM equation, although, in that case, the
phase $\Sigma(\xi,t)$ assumes a different expression. Solution
(\ref{gausson})-(\ref{fgausson}) describes a not dissipative
reversible process with a constant entropy
\begin{equation}
S^{\rm BG}=\sqrt{\pi\over\alpha}+\ln\sqrt{\pi\over\alpha}+\ln 2 \ .
\end{equation}

\sect{Conclusions}

In the stochastic quantization framework, we have approached the
problem of the quantization of a many-body classical system whose
kinetics is described by a general NFPE compatible with an arbitrary
trace form entropy. By a suitable modification of the original
Nelson method we have derived a family of NSEs describing, in the
mean field approximation, a system of collectively interacting
particles. The time evolution of the mean field $\psi({\bfm x},t)$
is given by means of a non-Hermitian operator $\mathrm{H}$ whose
nonlinearity, describing dissipative and damped effects, takes into
account the collective interactions among the many particles and the
background. This is a substantial different situation respect to the
canonical approach studied in
\cite{Scarfone1} which deals only with conservative systems.\\
We have investigated the existence of a special class of solutions
representing a localized object with particle-like properties
originated by the collective motion of the many particles of the
system. This solution describes a reversible and non dissipative
process
without production of entropy.\\
As an explicit example, we have studied the quantization of a
classical system governed by the BG entropy obtaining, in this way,
a NSE with a complex nonlinearity. This example emphasizes the
difference among the quantization method employed in the present
paper and the canonical one used in \cite{Scarfone1}. There,
starting from a classical system governed by the BG entropy, the
resulting quantum evolution equation is given by the linear
Schr\"odinger equation which describes the dynamics of an isolated
system. This is consistent with the widely accepted opinion that the
BG entropy governs systems with weakly and local interactions.
Differently, in the present case, although the particles in the
system must still be considered weakly interacting (due to the
presence of the BG entropy), there is a further interaction between
the system and the background which is accounted for the
nonlinearity in the evolution equation.

\vspace{10mm}
\noindent{\bf References}\\


\vfill\eject

\begin{thebibliography}{99}

\bibitem{Fermi} Fermi E 1955 Rend. Lincei {\bf5} 735

\bibitem{Ring} Ring P and Schuck P 1980 {\em The Nuclear Many-Body
Problem} (Springer, Berlin)

\bibitem{Negele} Negele J W and Orland H 1987 {\em Quantum Many-Body
Systems} (Addison-Wesley, Redwood)

\bibitem{Gross} Gross E P 1961 Nuovo Cimento {\bf20} 154; 1963 J. Math. Phys. {\bf4} 195

\bibitem{Pitaevskii} Pitaevskii L P 1961 Zh. Exsp. Teor. Fiz.
\bfm{50} 646; [1961 Sov. Phys. JETP {\bf13} 451]

\bibitem{Stringari} Stringari S 1996 Phys. Rev. Lett. {\bf77} 2360

\bibitem{Bialynicki} Bialynicki-Birula I and Mycielski J 1976 Ann.
Phys. {\bf100} 62

\bibitem{Weinberg} Weinberg S 1989 Ann. Phys. {\bf194} 336

\bibitem{Shimony} Shimony A 1979 Phys. Rev. A {\bf20} 394

\bibitem{Shull} Shull C G {\em et al.} 1980
Phys. Rev. Lett. {\bf44} 765

\bibitem{Bollinger} Bollinger J J {\em et al.} 1989 Phys. Rev. Lett. {\bf63} 1031

\bibitem{Hasse} Hasse R W 1980 Z. Phys. B {\bf37} 83

\bibitem{Hefter} Hefter E F 1985 Phys. Rev. A {\bf32} 1201

\bibitem{Bialynicki1} Bialynicki-Birula I and Mycielski J 1975
Commun. Math. Phys. {\bf44} 129

\bibitem{Bialynicki2} Bialynicki-Birula I and Madajczyk J L 1985
Phys. Lett. A {\bf108} 384

\bibitem{Maassen} Maassen H and Uffink J B M 1988 Phys. Rev. Lett.
{\bf60} 1103

\bibitem{Garbaczewski} Garbaczewski P 2006 J. Stat. Phys. {\bf123}
315

\bibitem{Tirnakli} Tirnakli U, Ozeren S F, Buyukkilic F and Demirhan D 1997 Z. Physik B {\bf104} 341

\bibitem{Olavo1} Olavo L S F, Bakuzis A F and Amilcar R Q 1999 Physica A {\bf271} 303

\bibitem{Rossani} Rossani A and Scrafone A M 2004 J. Phys. A: Math. Gen. {\bf37} 4955

\bibitem{Naudts} Czachor M and Naudts J 1999 Phys. Rev. E {\bf59}
R2497

\bibitem{Naudts1} Naudts J and der Straeten E V 2006 J. Stat. Mech.
P06015

\bibitem{Bender} Bender C M, Brody D C and Hook D W 2005 J. Phys. A:
Math. Gen. {\bf38} L607

\bibitem{Scarfone1} Scarfone A M 2005 Phys. Rev. E {\bf71} 051103; 2005 Rep. Math. Phys. {\bf55} 169

\bibitem{Nelson1} Nelson E 1966 Phys. Rev. {\bf150} 1079

\bibitem{Nelson2} Nelson E 1967 {\em Dynamical Theories of Brownian Motion} (Princeton University
Press, Princeton)

\bibitem{Nelson3} Nelson E 1985 {\em Quantum Fluctuations} (Princeton
University Press, Princeton)

\bibitem{Doebner} H.-D. Doebner and G.A. Goldin, Phys.
Lett. \bfm{A162}, 397 (1992); Phys. Rev. A \bfm{54}, 3764 (1996).

\bibitem{Scarfone0} Kaniadakis G, Quarati P and Scarfone A M 1998
Phys. Rev. E {\bf58} 5574

\bibitem{Scarfone7} Kaniadakis G and Scarfone A M 2003
Rep. Math. Phys. {\bf51} 225

\bibitem{Wallstrom} Wallstrom T C 1994 Phys. Rev. A {\bf49} 1613

\bibitem{Frank1} Frank T D 2005 {\em Nonlinear Fokker-Planck Equations: Fundamentals and Applications}
(Springer Series in Synergetics, Berlin)

\bibitem{Scarfone2} Kaniadakis G and Scarfone A M 2002 J. Phys. A:
Math. Gen. {\bf35} 1943

\bibitem{Scarfone3} Scarfone A M 2006 J. Phys. A: Math. Gen. {\bf38} 7037

\bibitem{Davison2} Davison M 1979 J.
Math. Phys. {\bf20} 1865; 1981 Lett. Math. phys. {\bf5} 523

\bibitem{Wang} Wang M S 1997 Phys. Rev. Lett. {\bf79} 3319

\bibitem{Bacciagaluppi} Bacciagaluppi G 1999 Found. Phys. Lett.
{\bf12} 1

\bibitem{Petroni} Petroni N C {\em et al.} 2003 Phys. Rev. ST Accel.
Beams {\bf 6} 034206

\bibitem{Davison1} Davison M 1979 Lett. Math. Phys. {\bf3} 271

\bibitem{Guerra} Guerra F and Morato L M 1983 Phys. Rev. D {\bf27}
1774

\bibitem{Kaniadakis} Kaniadakis G 2001 Physica A {\bf296} 405

\bibitem{Chavanis1} Chavanis P -H 2003 Phys. Rev. E {\bf68} 036108

\bibitem{Chavanis2} Chavanis P -H, Lauren\c{c}ot P and Lemou M
2004 Physica A {\bf341} 145

\bibitem{Bohm} Bohm D 1952 Phys. Rev. {\bf85} 166; 1952 {\bf85}
180

\bibitem{Madelung} Madelung E 1926 Z. Physik {\bf40} 332

\bibitem{Yasue} Yasue K 1981 J. Funct. Anal. {\bf41} 327

\end{thebibliography}
\end{document}